\documentclass[pre,twocolumn,showpacs]{revtex4}

\usepackage{natbib}
\usepackage{amssymb}
\usepackage{mathrsfs}
\usepackage{amsfonts}
\usepackage{graphicx}

\begin{document}

\title{First passage time of $N$ excluded volume particles on a line}

\author{Igor M. Sokolov}
\email{igor.sokolov@physik.hu-berlin.de}
\affiliation{Institut f{\"u}r Physik, Humboldt Universit{\"a}t zu Berlin,
Newtonstra{\ss}e 15, 12489 Berlin, FRG}
\author{Ralf Metzler}
\email{metz@nordita.dk}
\affiliation{NORDITA -- Nordic Institute for Theoretical Physics,
Blegdamsvej 17, 2100 Copenhagen \O, Denmark}
\author{Kiran Pant}
\affiliation{Department of Physics, Northeastern University, 111 Dana
Research Center, Boston, MA 02115, USA}
\author{Mark C. Williams}
\email{mark@neu.edu}
\affiliation{Department of Physics, Northeastern University, 111 Dana
Research Center, Boston, MA 02115, USA}

\begin{abstract}
Motivated by recent single molecule studies of proteins sliding on a DNA
molecule, we explore the targeting dynamics of $N$ particles ("proteins")
sliding diffusively along a line ("DNA") in search of their target site
(specific target sequence). At lower particle densities, one observes an
expected reduction of the mean first passage time proportional to $N^{-2}$,
with corrections at higher concentrations. We explicitly take adsorption and
desorption effects, to and from the DNA, into account. For this general case,
we also consider finite size effects, when the continuum approximation based
on the number density of particles, breaks down. Moreover,
we address the first passage time problem of a tagged particle diffusing among
other particles.
\end{abstract}


\date{\today}
\maketitle

\section{Introduction}

DNA-binding proteins can either be bound specifically, i.e., such that the
structure of the bound proteins exactly matches the entire DNA sequence it
covers, involving Gibbs free energies of some 10 kcal/mol and above; or it
can be bound non-specifically with lower Gibbs free energies. Non-specific
binding occurs when the bound protein matches only part of the covered DNA
sequence. A recent study showed that the repressor protein in
$\lambda$-infected E.coli bacteria is bound non-specifically with a Gibbs
free energy of some 4 kcal/mol, causing under typical conditions nearly
90 per cent of the repressor proteins to be bound non-specifically \cite{audun}.
In such a weak binding state, the protein can slide along the DNA, performing
a 1D diffusion process.

One of the primary tasks of DNA-binding proteins is the regulation of gene
expression, i.e., to determine whether (or not) a certain gene on the genome
is going to be transcribed by RNA polymerase. Having such processes in mind,
we refer to these binding proteins as transcription factors (TFs) in what
follows. The typical target search time of such a TF has received
renewed attention \cite{slutsky,coppey,marko,gerland}, after the detailed
investigations by Berg and von Hippel \cite{hippel}.
One-dimensional sliding motion of DNA-binding proteins along the DNA molecule
is an important ingredient in addition to three-dimensional volume diffusion
in the efficient specific target search that is observed in experiments
\cite{delbrueck,eigen,hippel}. There exist, however, situations when the
complete target search process of DNA-binding proteins occurs while being
non-specifically attached to the DNA molecule, i.e., without detaching from
the DNA before hitting the target. This could be recently proved for
bacteriophage T4 single-stranded DNA binding protein gp32 \cite{pant,pant_a}.

Gene regulation is a highly relevant example of a first passage time
process, that can, in addition, be probed experimentally on the single
molecule level. While one usually considers the first passage of a single
random walker, or an ensemble of phantom random walkers, the sliding
proteins on the DNA are clearly mutually excluding. To understand their
target search quantitatively, one needs a theoretical model for the
first passage of non-phantom particles. Surprisingly, there have been
studied only a few cases of diffusion processes of mutually excluding
particles, for instance, the diffusivity of particles on a line
\cite{aslangul}. It should be noted that while
some of the results below are known per se for the case of one-particle
diffusion or for phantom particles \cite{redner,rice}, in the present case they
are based on a mapping of the case of impenetrable particles, a problem
that, to our knowledge, has not been studied so far.
We also note that the problem pursued here is therefore also of a more
generic interest, pertaining to the modeling of charge carrier motion
in effectively one-dimensional geometries (nanowires, etc.) or traffic
flow, among others.

In what follows, we establish a theory for the first passage dynamics of
mutually excluding particles along a line ("DNA"). We explicitly take
adsorption of
particles to and desorption from the DNA into account, mimicking
possible volume excursions of the proteins. Apart from the dilute case,
we also address the dense case and the possibility of having more than
one species of particles. Our analytical findings are corroborated by
simulations.

\section{Scaling approach}

In the simplest case when $N$ identical, mutually excluding particles of size
$\lambda$ diffuse along a line of length $L$, we can obtain insight into
the associated first passage process from scaling arguments. To be precise,
the first passage is considered for a target placed at the origin ($x=0$)
for particles, that are initially randomly distributed along the line $L$.
We first address the dilute case when the length $N\lambda$ occupied by the
sliding particles can be neglected in comparison to the length $L$. Finite
size effects are regarded at the end of this section.

\begin{figure}
\begin{center}
\includegraphics[width=6.2cm,angle=270]{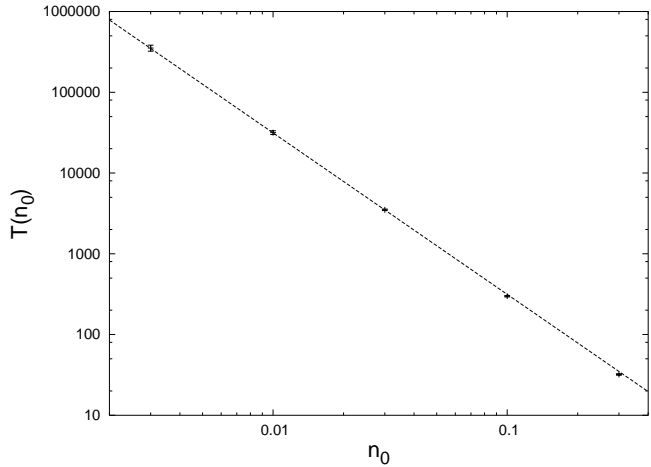}
\end{center}
\caption{Mean first passage time $T(n_0)$ in a one-sided system, as function
of the density $n_0$ of excluding walkers that cannot occupy the same
lattice site. The target is placed at $x=0$. The maximum density is $n_0=30$\%,
whereby each particle occupies one lattice site. The dashed line corresponds to
the exact result $T(n_0)=\pi/n_0^2$ from equation (\protect\ref{tau}), with the
dimensionless diffusion coefficient $D_{\mathrm{1d}}=1/2$. The simulation
data agree nicely with the dilute limit, with only a slight deviation for
larger densities. Each data point corresponds to $10^5$
runs, except for $10^3$ realizations for the lowest density. Note the
comparatively small error bars.}
\label{targ}
\end{figure}

On average, it takes a random walker the time $T\simeq L^2/D_{\mathrm{1d}}$
to cover the distance $L$ by unbiased diffusion. Here, $D_{\mathrm{1d}}$ is the
diffusion coefficient for 1D motion on a line, and the symbol $\simeq$
indicates that we neglect constant prefactors. If there are $N$ identical
particles placed randomly over the line $L$, they are separated by an average
length $L/N$, i.e., each of them has a free diffusion length $L/N$. For the
first of these particle to hit the target site, this requires a characteristic
target search time
\begin{equation}  \label{tdil}
T_{\mathrm{dil}}(N)\simeq\frac{L^2}{D_{\mathrm{1d}}N^2}=\frac{1}{D_{\mathrm{%
1d}}n_0^2},
\end{equation}
with $n_0=N/L$ being the concentration of particles. The index is meant to
distinguish the dilute result from the result (\ref{tfull}) when finite size
effects come into play.

We performed a simulation of particles on a line during which each particle
attempts a jump to its left or right nearest neighbor lattice point per unit
time. In case the corresponding site is occupied, the step is forbidden, and
the particle remains at its original site. The associated (dimensionless)
diffusion coefficient of a \emph{single} particle per unit length square and
unit time is $D_{\mathrm{1d}}=1/2$. Figure \ref{targ} shows the results for
the mean target search time $T(n_0)$ of this simulation in dependence of
the density $n_0$ of particles. We find nice agreement with the expected
inverse square dependence of $T(n_0)$ on the density $n_0$. The line through
the data points corresponds to the analytical result from Eq.~(\ref{tdil})
with a prefactor given by Eq.~(\ref{tau}) without adjustable parameters. 
The results demonstrate that the theoretical
approximation leading to the $1/N^2$ behavior remains reasonable even at
rather high concentrations, at which the interparticle distance becomes of
the order of the step lengths. In the next section, we derive the $1/N^2$
scaling analytically in a continuum approximation.

\begin{figure}
\begin{center}
\includegraphics[width=6.2cm,angle=270]{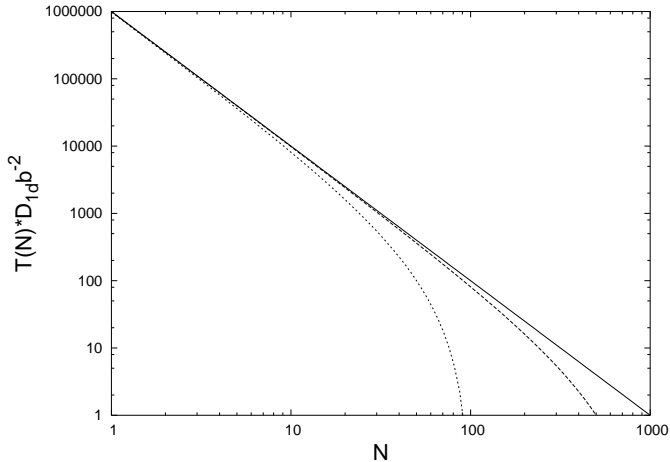}
\end{center}
\caption{Behavior of the mean first passage time $T(N)$ as a function of
the number $N$ of TFs attached to a DNA of length $1000b$ according to
equation (\ref{tfull}), for the dilute
case (---), TF-size $\lambda=b$ (long-dashed) and $\lambda=10b$
(short-dashed). Excluded volume effects reduce the target search time
$T(N)$.}
\label{targ1}
\end{figure}

Experimentally, for instance in in vivo studies of proteins binding to a
DNA molecule, the diluteness condition is perfectly adequate, compare, for
instance, Ref.~\cite{mark}. By increasing the protein concentration or their
binding strength through different ambient salt conditions, the concentration
of bound proteins can be increased such that finite size effects indeed come
into play. Similarly, the presence of many different species of proteins
leads to a rather crowded DNA molecule. Similar considerations apply, of
course, to other systems. Defining the occupation ratio
\begin{equation}
f=\frac{N\lambda}{L},
\end{equation}
we can express the diluteness condition through $f \ll 1$. To include finite
size effects when this limit is not fulfilled in our scaling approach, we
only need to consider the reduced length of the line available to the random
walking particles. This reduced length is
$L_{\mathrm{red}}=L-N\lambda$, so that we obtain 
\begin{equation}  \label{tfull}
T(N)\simeq\frac{(L-N\lambda)^2}{D_{\mathrm{1d}}N^2}=T_{\mathrm{dil}}(N)
(1-f)^2,
\end{equation}
for the scaling of the mean target search time with the number $N$ of
particles. Figure \ref{targ1} compares the dilute $1/N^2$ scaling with
the finite size effects predicted by the excluded volume expression
(\ref{tfull}).

\section{The continuum approximation}

In this section, we verify the above scaling result for the dilute case,
$T(N)\simeq L^2/\left(D_{\mathrm{1d}}N^2\right)$, through an analytic
treatment in the continuum approximation, replacing the individual TFs
through the particle density $n(x,t)$. In addition, we include
explicitly adsorption and desorption effects with rates $k_0$ and $k_1$.

To be able to take the continuum limit, we consider large systems (long
DNA) with many ($N\gg 1$) searching TFs, such that the concentration of TFs
on the DNA is much smaller than unity; that is, $f\ll 1$. In other words, the
diffusion time through the whole system, $T_1\simeq L^{2}/D_{\mathrm{1d}}$,
is much larger than the typical first passage time corresponding to the
characteristic target search time, being of the order of $T\simeq 1/(f^2D_{
\mathrm{1d}})$. We note that for $f\ll 1$, the fraction $f$ depends linearly
on the volume concentration $C$ of TFs, according to the McGhee and von Hippel
isotherm \cite{ghee}.

We start by considering a one-sided problem (one target site at $x=0$ of a
semi-infinite DNA). The time evolution of the number concentration $n(x,t)$
at position $x$ at time $t$ on the semi-infinite interval is then given by the
diffusion-reaction equation 
\begin{equation}
\label{rde}
\frac{\partial n}{\partial t}=D_{\mathrm{1d}}\frac{\partial^2}{\partial x^2}%
n -k_1n+k_0.
\end{equation}
Apart from diffusion, in this equation we take into account adsorption (with
rate $k_0$) and desorption (with rate $k_1$) of the TFs, where the desorption
is proportional to the number concentration of TFs on the DNA. Apart from
real physical absorption/desorption processes, this approach might mimic
other nonlocal processes such as macrohops (3D volume sojourns) and
intersegmental transfer (hopping from one segment of the DNA to another,
chemically remote segment, that is close by in geometric space due to
looping of the DNA) in a mean field sense. Following
Smoluchowski's approach to diffusion-controlled reactions, we represent the
target site by an absorbing boundary condition at $x=0$, i.e., when a
diffusing particle hits this site, it will be removed. The possibility of
double occupation of sites is disregarded, as it represents a higher order
effect proportional to $f^2$. Moreover, the fact that particles are
impenetrable to each other does not change the behavior at low
concentrations, since, neglecting the excluded volume, on encounter of two
particles it does not matter whether the right particle always stays to the
right of the other particle (impenetrable particles), or whether they change
roles and the right particle becomes the left one (phantom particles), as
long as the particles are indistinguishable, in contrast to the case of
distinguishable particles addressed below. Finite size effects due to high
occupation, violating the diluteness condition $f\ll 1$ will also be addressed
below.

Finding the target corresponds to the event when the first particle hits the
target site. Mathematically, this is equivalent to the first passage time of a
particle from a site $x>0$ to $x=0$, given by the particle flux into the
reaction center, $j(t)=D_{\mathrm{1d}}\left.\partial n/\partial x\right|_{
x=0}$. The survival probability $\mathscr{S}(t)$ of the target site (i.e.,
the probability of not yet having been hit by a TF, not to be confused with
the survival of the particles along the DNA) is consequently given by the
first-order kinetic equation 
\begin{equation}
\label{kinetic}
\frac{d}{dt}\mathscr{S}(t)=-j(t)\mathscr{S}(t).
\end{equation}
The change of the survival probability, of not having been hit, of the
target site is thus the product of the probability of not having been hit
previously times the magnitude of the influx of particles.
The formal solution of Eq.~(\ref{kinetic}) reads 
\begin{equation}
\label{surv}
\mathscr{S}(t)=\exp\left(-\int_0^tj(t^{\prime})dt^{\prime}\right).
\end{equation}
In what follows we use the notation $J(t)=\int_0^tj(t^{\prime})dt^{\prime}$.
The first passage time \emph{density\/} is then given by 
\begin{equation}
\psi (t)=-\frac{d}{dt}\mathscr{S}(t)=j(t)\exp\left(-J(t)\right) .
\end{equation}
In our one-sided problem, the mean first passage time becomes 
$T=\int_0^{\infty} t\psi(t)dt=-\int_{0}^{\infty }t[d\mathscr{S}(t)/dt]dt$, i.e., 
\begin{equation}
T=\int_0^\infty \mathscr{S}(t^{\prime})dt^{\prime}.
\end{equation}

To obtain an explicit expression for $\mathscr{S}(t)$, we solve the reaction-diffusion
equation (\ref{rde}) by Laplace transformation techniques. With the initial
condition $n(x,0)=n_{0}\Theta (x)$, where $\Theta(x)$ is the Heaviside jump
function, we obtain for all $x>0$ for the Laplace transform $\tilde{n}(x,u)$: 
\begin{equation}
u\tilde{n}-n_0=D_{\mathrm{1d}}\frac{\partial^2}{\partial x^2}\tilde{n}+ 
\frac{k_0}{u}-k_1\tilde{n},
\end{equation}
i.e., a linear inhomogeneous differential equation of the form 
\begin{equation}
\tilde{n}^{\prime\prime}-\Lambda \tilde{n} + B=0  \label{LapTran}
\end{equation}
with $\Lambda=(k_1+u)/D > 0$ and $B=(k_0/u+n_0)/D >0$. The boundary
conditions we impose are of the absorbing Dirichlet type $n(0,u)=0$ at the
target site placed at the origin, and the natural boundary condition
$n(x,u)<\infty $
for $x\rightarrow \infty $. The corresponding solution reads 
\begin{equation}
\tilde{n}(x,u)=\frac{k_{0}+un_{0}}{u\left( k_{1}+u\right) }\left( 1- e^{-x%
\sqrt{(k_1+u)/D_{\mathrm{1d}}}}\right) .
\end{equation}

From this expression, we find for the flux $j(t)$ in Laplace space 
\begin{equation}
\tilde{j}(u)=D_{\mathrm{1d}}\left. \frac{\partial \tilde{n}(x,u)}{\partial x}
\right|_{x=0}=\sqrt{D_{\mathrm{1d}}}\frac{k_{0}+un_{0}}{u\sqrt{k_{1}+u}},
\label{jvonu}
\end{equation}
an expression whose inverse Laplace transform can be calculated explicitly,
yielding 
\begin{equation}
j(t)=\sqrt{D_{\mathrm{1d}}}\left[\frac{k_0}{\sqrt{k_1}}\mathrm{erf}\sqrt{k_1t%
} +n_0\frac{e^{-k_1t}}{\sqrt{\pi t}}\right].
\end{equation}
The survival probability of the target site then is given by $\mathscr{S}(t)=\exp
\left( -J(t)\right) $ with 
\begin{eqnarray}
J(t)=&&\sqrt{D_{\mathrm{1d}}}\left[\frac{k_0}{k_1}\left(t\sqrt{k_1} \mathrm{%
erf}\sqrt{k_1t}-\frac{\mathrm{erf}\sqrt{k_1t}}{2\sqrt{k_1}} \right.\right. 
\nonumber \\
&&\left.\left.\hspace{1.2cm}+\frac{\sqrt{t}}{\sqrt{\pi}}e^{-k_1t}\right) +n_0%
\frac{\mathrm{erf}\sqrt{k_1t}}{\sqrt{k_1}}\right].  \label{exponent}
\end{eqnarray}

Without adsorption and desorption (i.e., $k_{0}=k_{1}=0$), we obtain the
survival provability 
\begin{equation}
\mathscr{S}(t)=\exp \left( -2n_{0}\sqrt{\frac{D_{\mathrm{1d}}t}{\pi }}\right)
\label{Pnoad}
\end{equation}
and first passage time density 
\begin{equation}
\psi(t)=\frac{n_0\sqrt{D_{\mathrm{1d}}}}{\sqrt{\pi t}}\exp\left(-2n_0\sqrt{%
\frac{D_{\mathrm{1d}}t}{\pi}}\right).
\end{equation}
We thus find for the mean first passage time $T=\int_0^{\infty }\mathscr{S}(t)dt$ the
simple form 
\begin{equation}
T_{\mathrm{line}}=\frac{\pi}{2}\frac{1}{n_0^2D_{\mathrm{1d}}}  \label{tau}
\end{equation}
showing the typical $n_0^{-2}$ dependence on the initial concentration.

The survival provability for the general case with non-vanishing
rates $k_0$ and $k_1$ becomes 
\begin{eqnarray}
\mathscr{S}(t)=&&\exp\left[-\sqrt{D_{\mathrm{1d}}}\left(k_0k_1t-k_0/2+n_0k_1\right) 
\frac{\mathrm{erf}\sqrt{k_1t}}{k_1^{3/2}}\right.  \nonumber \\
&&\hspace{1.2cm}\left.-\frac{k_0}{k_1} \frac{\sqrt{D_{\mathrm{1d}}t}}{\sqrt{%
\pi}}\exp(-k_1t)\right].
\label{surv_general}
\end{eqnarray}
Eventually (for $t\gg k_1$), an exponential decay $\sim\exp\left(-\sqrt{D_{
\mathrm{1d}}}k_0k_1t\right)$ is reached. From this asymptotic behaviour, we
can deduce the approximate dependence $T\approx\left(\sqrt{D_{\mathrm{1d}}}
k_0k_1\right)^{-1}$. As the adsorption rate $k_0$ is proportional to the
concentration $C$ of TFs in volume, we obtain the typical $T\sim C^{-1}$
dependence of the mean target search time under volume exchange conditions.
This contrasts the $T\sim n_0^{-2}$ behaviour for 1D sliding exchange found
in Eq.~(\ref{tau}). Given that $n_0\simeq C$ for $n_0\ll 1$, the latter 
corresponds to the $T\simeq C^{-2}$ scaling demonstrated in Fig.~\ref{rates}
below. In general, there will be a combination of both behaviours, depending
on the values of the various system parameters.

In the case of no adsorption $k_0=0$ but non-vanishing desorption $k_1\neq 0$
that corresponds to a situation with vanishing concentration of TFs in the
free volume, the function 
\begin{equation}
J(t)=\sqrt{D_{\mathrm{1d}}}n_0\frac{\mathrm{erf}\sqrt{k_1t}}{k_1^{1/2}}
\end{equation}
is bounded from above, by $n_0\sqrt{D_{\mathrm{1d}}/k_1}$, and the survival
probability $\mathscr{S}(t)$ never reaches zero (all particles desorb with a nonzero
probability without ever reaching the target site $x=0$), and the
probability density $\psi (t)$ is a non-proper one, corresponding to a
diverging mean first passage time. In all other cases $\psi(t)$ is a proper
probability density, and the mean target search time $T$ is finite.

Performing an expansion in powers of $t$ (the corresponding series contains
only the half-integer powers), we find for the function $J(t)$ in the general
case with finite $k_0$, $k_1$: 
\begin{eqnarray}
J(t)&=&\sqrt{\frac{D_{\mathrm{1d}}}{\pi}}\left[2n_0t^{1/2}+\frac{2}{3}k_1
\left(2\frac{k_0}{k_1}-n_0\right) t^{3/2}\right.\nonumber\\
&&\left.+\frac{1}{15}k_1^{2}
\left(-4\frac{k_0}{k_1}+3n_0\right) t^{5/2}+...\right] 
\label{expansion}
\end{eqnarray}
so that the $n$-th term of the expansion has a structure 
$k_1^{n-1}(a_nk_0/k_1 +b_nn_0)t^{(2n-1)/2}$. Thus, in essence, this expansion
corresponds to an expansion in powers of $k_1$. Note that $k_0/k_1=n_s$ is a
steady-state concentration of proteins in the absence of the absorbing
target site. As long as both $k_0$ and $k_1$ are small, the overall
behavior given by equation (\ref{tau}) is preserved, provided the initial
concentration $n_0$ is not too small.
In the case without desorption ($k_1 \rightarrow 0$) we get 
\begin{equation}
\mathscr{S}(t)=\exp \left( -2n_{0}\sqrt{\frac{D_{\mathrm{1d}}t}{\pi }}
-\frac{4}{3}\sqrt{\frac{D_{\mathrm{1d}}}{\pi}} k_0 t^{3/2}  \right).
\label{Padnode}
\end{equation}
This equation is important in what follows, when finite-size effects are
considered.

\begin{figure}
\begin{center}
\includegraphics[width=8.8cm]{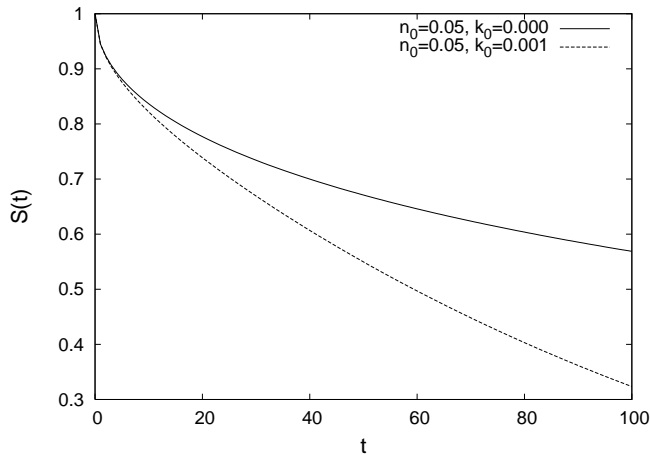}
\end{center}
\caption{Survival probability of the target site, i.e., the probability that
no TF has reached the specific binding site, according to Eqs.~(\ref{Pnoad})
and (\ref{Padnode}). This case corresponds to vanishing desorption, $k_1=0$.
The plot parameters are indicated in the figure.}
\label{survival}
\end{figure}

\begin{figure}
\begin{center}
\includegraphics[width=8.8cm]{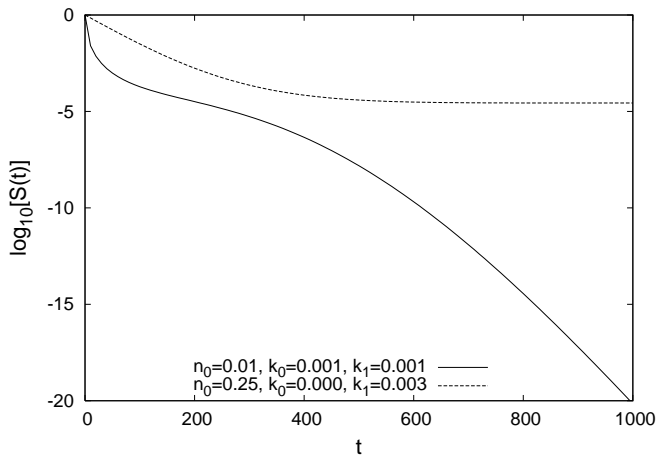}
\end{center}
\caption{Survival probability $\mathscr{S}(t)$ from Eq.~(\ref{surv_general})
for finite desorption rate $k_1$. Note the logarithmic ordinate; the plot
parameters are indicated in the figure. The incomplete decay in the case of
vanishing adsorption, $k_0=0$ is distinct.}
\label{survival1}
\end{figure}

In Fig.~\ref{survival}, we plot the survival probabilities from
Eqs.~(\ref{Pnoad}) and (\ref{Padnode}), for an initial line density
of TFs of $n_0=0.05$. Both cases correspond to vanishing desorption rate,
$k_1=0$, and therefore $\mathscr{S}(t)$ decays completely for large times.
This decay of the survival probability of the target site in both cases
follows the same behavior for short times, until the adsorption according
to Eq.~(\ref{Padnode}) leads to faster target search and therefore to a
quicker decay of $\mathscr{S}(t)$. Similarly, Fig.~\ref{survival1} shows
the survival probabilities in the general case corresponding to
Eq.(\ref{surv_general}); note the logarithmic ordinate. For vanishing
adsorption but finite desorption, the expected incomplete decay of
$\mathscr{S}(t)$ is observed, whereas for finite ad- and desorption the
transition between the different contributions in expression
(\ref{surv_general}) is visible, eventually approaching the simple
exponential pattern, that corresponds to a straight line in this plot.

The two-sided problem (a ring geometry with a perimeter that is much larger
than the typical interparticle distance) corresponds to the situation where
two competing processes occur, i.e., the survival probability of having an
empty target site changes in time through the influx of TFs from both sides.
This practically corresponds to using twice the probability current $j$ in
equation (\ref{kinetic}) due to symmetry, and therefore to 
\begin{equation}
\mathscr{S}(t)=\exp\Big(-2J(t)\Big)
\end{equation}
with $J(t)$ given by equation (\ref{exponent}). The corresponding mean first
passage time for the case $k_0=k_1=0$ is then given by 
\begin{equation}  \label{circ}
T_{\mathrm{ring}}=\frac{\pi}{8}\frac{1}{n_0^{2}D_{\mathrm{1d}}},
\end{equation}
that is by a factor of 4 smaller than in the one-sided case. Result (\ref
{circ}) is also confirmed by numerical simulations. We note that the
reduction by a factor 4 can be easily understood by mapping the circle with
one absorbing site onto a line whose both ends are absorbing boundaries. It
then corresponds to two one-sided geometries as considered above, but with an
effective length of $L/2$. With the definition of the initial number
concentration $n_0$, this reproduces the factor 4.

For direct comparison with the experimental data, figure \ref{rates} shows
an alternative way to present the numerical data from figure \ref{targ},
in dimensional form of the rate $k_a$ in units of 1/s versus the volume
protein concentration $C$ in units of M. For the conversion, we use the
relation $n_0=K_{\mathrm{ns}}\mu C$, with the nonspecific binding constant
$K_{\mathrm{ns}}=2.5\cdot 10^5$ M$^{-1}$, and the SSB binding size $\mu=7$
in units of nucleotides \cite{pant,pant1}. By logarithmic least
squares fit to the shown data measured at 100 mM salt, we obtain for the
dimensional diffusion constant $D_{\mathrm{1d}}$
of 1D sliding along the dsDNA the value $D_{\mathrm{1d}}=3.3\cdot 10^{-9}
\mathrm{cm}^2/\mathrm{sec}$, that is nicely within the reported range
$10^{-8}\ldots 10^{-9}\mathrm{cm}^2/\mathrm{sec}$ for this salt concentration
\cite{pant,pant1}. This corroborates the validity of our rather simple
analytical model for the target search of a truncate of the gp32 protein.
Note that the experimental situation with two
target sites at either end of the DNA
molecule corresponds to the result (\ref{circ}).

\begin{figure}
\begin{center}
\includegraphics[width=6.2cm,angle=270]{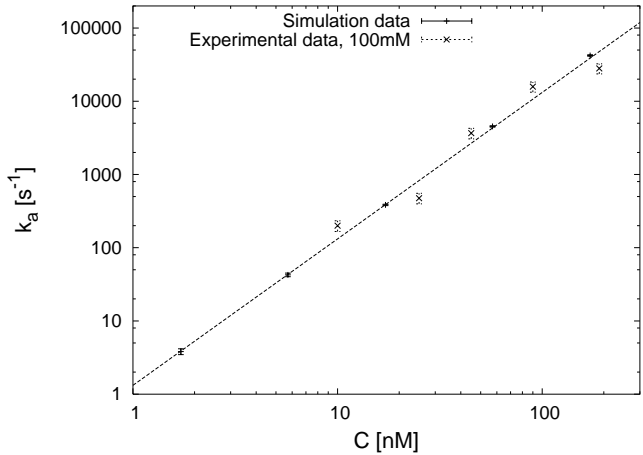}
\end{center}
\caption{Dimensional binding rate $k_a$ in 1/s as function of protein
concentration $C$ in nM, converted from figure \protect\ref{targ} for
parameters corresponding to 100 mM salt. The fitted 1D diffusion constant
for sliding along the dsDNA is $D_{\mathrm{1d}}=3.3\cdot 10^{-9}\mathrm{cm}^2/
\mathrm{sec}$, located nicely within the experimental value $10^{-8}\ldots
10^{-9}\mathrm{cm}^2/\mathrm{sec}$ \protect\cite{pant}.}
\label{rates}
\end{figure}

\section{Finite-size effects}

In the previous section we discussed the case of a semi-infinite DNA, and
argued that the case when the target site is situated somewhere in
the middle of the molecule, can be inferred from that result.

Now we consider the finite-size situation
(again one-sided), with a target site situated on one side of a chain,
and with another side closed by a ``stopper'', for instance, a polystyrene
bead in an optical tweezers setup, such that the sliding proteins observe
a reflecting boundary condition. This consideration is necessary
to discuss finite-size effects, and also to derive results explicitly used
in the next section. The situation where there are two portions of the chain
to the left and to the right from the target site corresponds to two
independent reaction channels, so that the mean reaction time follows from
those in the left and in the right intervals: $1/T=1/T_L+1/T_R$.

\begin{figure}
\begin{center}
\includegraphics[width=6.2cm,angle=270]{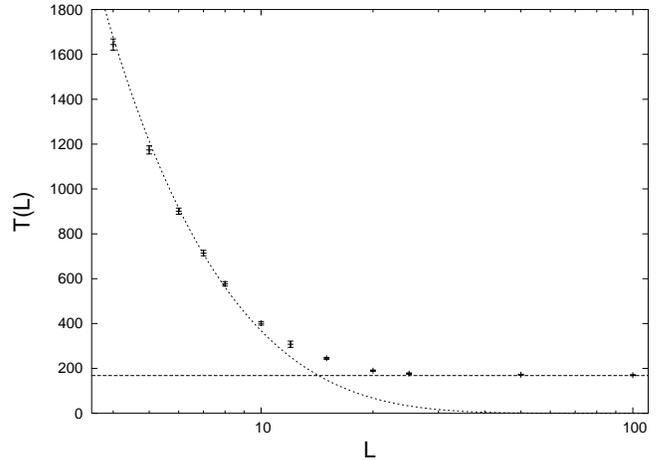}
\end{center}
\caption{Mean first passage time $T(L)$ of target search in a pronouncedly
finite system of length $L$, corresponding to the one-sided system with
target site at $x=0$ and reflecting boundary condition at $x=L$. We chose
the parameters $D_{\mathrm{1d}}=1/2$ for the dimensionless diffusion
constant, the initial protein density $n_0=10$\%, the adsorption rate $k_0=10^{-4}$,
and vanishing desorption rate $k_1=0$. The curved dashed line corresponds
to equation (\protect\ref{Tlim}), that approximates small systems, while
the horizontal dashed line at $T=168$ is determined by numerical integration
of $\mathscr{S}(t)$, equation (\ref{Padnode}).
Each data point represents $10^5$ runs. Again, note the rather small error bars.}
\label{fin_eff}
\end{figure}

To consider this situation on an interval of length $L$ with exactly $N$
TFs, we have to solve our equation (\ref{rde}) with the boundary conditions $%
n(0,t)=0$ (reacting center), $n^{\prime }(L,t)=0$ (blocked end), and with
initial condition $n(x,0)=n_{0}=N/L$. Under Laplace-transformation this leads
us to equation (\ref{LapTran}), now with corresponding boundary conditions. The
solution then becomes
\begin{equation}
\tilde{n}(x,u)=\frac{b}{\lambda }\left( 1-\cosh x\sqrt{\lambda }+\tanh L%
\sqrt{\lambda }\sinh x\sqrt{\lambda }\right),
\label{concentr}
\end{equation}
so that the Laplace transform of the probability current reads: 
\begin{equation}
\tilde{j}(u)=\sqrt{D_{\mathrm{1d}}}\frac{k_{0}+un_{0}}{u\sqrt{k_{1}+u}}\tanh 
\frac{L\sqrt{k_{1}+u}}{\sqrt{D_{\mathrm{1d}}}}.
\end{equation}
This expression tends to our equation (\ref{jvonu}) in the limiting case $%
L\rightarrow \infty $. In the general case there does not exist
a closed expression for $j(t)$ and thus for $T$. However, for small
enough $L$ ($L\sqrt{k_{1}}/\sqrt{D_{\mathrm{1d}}}\ll 1$, i.e. in the case when
the diffusion time along the $L$-interval is so small, that practically no
desorption takes place) we can approximate $\tanh x$ by the value of its
argument and obtain 
\begin{equation}
\tilde{j}(u)=\frac{L}{u}(k_{0}+n_{0}u).
\end{equation}
This result implies
\begin{equation}
j(t)\simeq \left[ k_{0}+n_{0}u\right] L,
\end{equation}
and
\begin{equation}
J(t)\simeq \left[ k_{0}+n_{0}\delta (t)\right] L,
\end{equation}
so that we find the survival probability 
\begin{equation}
\mathscr{S}(t)\simeq e^{-n_{0}L}e^{-k_{0}Lt}.  \label{Surv1}
\end{equation}
The latter result leads to the approximate form
\begin{equation}
T\simeq \frac{e^{-n_{0}L}}{k_{0}L}.  \label{Tlim}
\end{equation}
This behavior will be of importance in what follows. In figure \ref{fin_eff},
we show results from simulations on a finite system, demonstrating that the
predicted asymptotic behavior, equation (\ref{Tlim}), in fact describes the
behavior of the system quite accurately for smaller $L$, and eventually
reaches a constant value for larger system sizes (note that the density
$n_0$ of proteins is kept constant).

The enumerator in equation (\ref{Tlim})
is the probability that no TFs are initially present in the interval. If
there were any, the target finding could typically occur within the time
interval $\tau \simeq L^{2}/D_{\mathrm{1d}}$ which is extremely small if $D_{\mathrm{1d}}$ is small
enough. However, there is a nonzero probability (equal to $e^{-n_{0}L}$)
that no TFs are initially found in the interval. In this case one has to
wait on the average $1/k_{0}L$ until a TF is adsorbed, a slow process which
governs the overall expression kinetics. This is the true asymptotics of the
waiting time in the case when 
$k_0$ is very small, so that the absorption time $1/k_{0}L$ is much larger than the
typical diffusion time over the interval $L$ being of the order of $T_D \simeq L^2/D_{\mathrm{1d}}$.   

We note that the situation considered here is pertinent to the
grand canonical ensemble ($N$ fluctuates around the mean value 
$N=n_0 L$); the canonical situation ($N$ fixed) is discussed in the Appendix.

\section{Different species of transcription factors}

The picture changes if we regard TFs of different species. If the relative
concentration of ``relevant'' TFs is high enough, the situation stays
practically the same as before, since the ``dummy'' proteins simply act as the
effective ``boundaries'' reducing the length of the search region to $%
\tilde{L}=L/N_{\mathrm{dummy}}$ around the target site. This simple
assumption is realistic since specifically bound TFs would, for most practical
purposes, represent immobile barriers (the Gibbs free energy for specific
binding is larger than for non-specific binding). Since,
however, the effective search time depends only on the overall
concentration of relevant proteins, the typical search time will not change
considerably, unless the situation occurs that no relevant proteins are
encountered within the search region with appreciable probability. This
situation
takes place if the concentration of dummy TFs gets of the order of or larger
than the concentration of relevant TFs. The reaction can take place only if a
relevant particle is situated in the same interval between the
barriers as the target site is. The mean waiting time in this case can
be obtained from the result of the previous section. Let us consider the
interval to one side of the target site. The length of this interval
be $L$. Assuming independent positions of all TFs, we can obtain the joint
probability distribution of the length of the interval between the reaction
center and the next boundary protein, and of the mean initial concentration
of relevant TFs inside, $p(n_{0},L)$ (note here that the variable $N$ is
discrete, while $L$ is continuous). Noting that the actual initial
concentration $n_{0}$ in each realization is $n_{0}=N/L$ and that the
density of the waiting time distribution for a given (non-fluctuating) $n_{0}
$ and $L$ is given by a function $\psi (t;L,n_{0},k_{0},k_{1})$, the overall
waiting time distribution yields as a mixture, i.e., by simple averaging 
\begin{equation}
\Psi (t)=\int_{0}^{\infty }dLp(L)\psi (t;L,n_{0},k_{0},k_{1}),
\end{equation}
where $p(L)$ is the probability density to find a specifically bound TF at
the distance $L$ from the target site. The corresponding mean waiting
time in the one-sided problem follows then as 
\begin{equation}
\bar{T}=\int_{0}^{\infty }dLp(L)T(L,N/L,k_{0},k_{1})
\end{equation}
with the weight $T(L,n_{0},k_{0},k_{1})=\int_{0}^{\infty }t\psi
(t;L,n_{0},k_{0},k_{1})dt=\int_{0}^{\infty }\mathscr{S}(t;L,n_{0},k_{0},k_{1})dt$, where
$\psi $ represents the waiting time probability density function,
and $\mathscr{S}(t)=\int_{t}^{\infty }\psi
(t^{\prime })dt^{\prime }$ being the survival probability. Assuming
Poissonian statistics of the distribution of TFs we find that the
distribution of $L$ is exponential, $p(L)=c_{s}e^{-c_{s}L}$, with $c_{s}$
being the concentration of specifically bound TFs. Here a clear difference
between the one-sided and the two-sided problem emerges. 

We give here the explicit results only for the case when $k_0 < c_s^3 D$.
The mean waiting
time in a one-sided problem is then given by equation (\ref{Tlim}). Averaging this
expression over the distribution of the lengths of the intervals between the
target site and the blocking specifically bound TF we see that the
corresponding expression 
\begin{equation}
\bar{T}_1 =\frac{c_{s}}{k_{0}}\int_{0}^{\infty }\frac{1}{L}e^{-(n_{0}+c_{s})L}dL
\end{equation}
diverges. This divergence has to do with the possibility of immediate
blocking, which gets evident when we return to the initial, discrete
situation: since it is possible that a specifically bound TF is an immediate
neighbor of the target site, the reaction is simply impossible. Of
course, one can overcome this difficulty by assuming that there exists a
minimal size of such an interval  $L_{\min }$ (or that the immediate
absorption of a relevant TF on the center is possible, which, from the
mathematical point of view, is equivalent to putting this minimal length
equal to a size of the target site). Assuming this $L_{\min }$ to be
small compared to all other spatial scales of the problem so that 
$p(L)\simeq c_{s}e^{-c_{s}L}/e^{-c_{s}L_{\min }}$, we obtain asymptotically 
\begin{eqnarray}
\bar{T} & = & \frac{c_{s}}{k_{0}e^{-c_{s}L_{\min }}}\int_{L_{\min }}^{\infty }
\frac{1}{L}e^{-(n_{0}+c_{s})L}dL \nonumber \\
& = & \frac{c_{s}}{k_{0}e^{-c_{s}L_{\min }}}\Gamma \left[ 0,(n_{0}+c_{s})L_{\min }\right]
\end{eqnarray}
($\Gamma (x,y)$ being the incomplete $\Gamma $-function) which grows
very slowly (logarithmically) for $L_{\min }\rightarrow 0$. 

Let us now turn to the two-sided situation. In this case the survival
probability of the target site $\Psi
(t)=\mathscr{S}(t;L_{1},n_{0},k_{0},k_{1})\mathscr{S}(t;L_{2},n_{0},k_{0},k_{1})$ where $L_{1}$
and $L_{2}$ are the lengths of free intervals to the left and to the right
from the target site: it survives up to time $t$ if the TF comes to it
neither from the right nor from the left. Using equation (\ref{Surv1}) valid
for $k_{1}$ small we get
\begin{eqnarray}
\bar{T}_{2} & = & \int_{0}^{\infty }dt
\int_{0}^{\infty }\int_{0}^{\infty}dL_{1}dL_{2}c_{s}^{2}
e^{-(c_{s}+n_{0}+k_{0}t)(L_{1}+L_{2})} \nonumber \\
& = & c_{s}^{2}\int_{0}^{\infty }\frac{dt}{(c_{s}+n_{0}+k_{0}t)^{2}} \nonumber \\
&=&\frac{c_{s}^{2}}{(c_{s}+n_{0})k_{0}}.
\end{eqnarray}
In this case the mean waiting time (still fully defined by the absorption)
is finite, since the probability that the target site is blocked from
both sides is negligibly small.

\section{Conclusions}

We derived analytically the first passage time behavior for a set of $N$
mutually excluding particles on a line in the dilute limit. As predicted
from scaling arguments, the corresponding mean first passage time decays
inversely with the square of the number of particles. The analytical
behavior was corroborated by simulations results, showing nice agreement
without a free parameter. Comparison with experimental results from the
one-dimensional target search of the bacteriophage T4 protein gp32 produces
a very reasonable fitted one-dimensional diffusion constant of the sliding
protein.

Having in mind the target search of transcription factors on a long DNA,
during which one-dimensional sliding motion along the DNA is interrupted
by three-dimensional volume excursions, we included desorption from and
adsorption to the DNA. These affect the time evolution of the survival
probability of the specific target sequence, that may be of importance
to the design of related in vitro experiments. Moreover, the obtained
description may be relevant to other (bio)chemical systems as well as
nano-setups, for instance, the one-dimensional diffusion of particles
in a nano-channel, and their escape through a T-junction.

Finally, we discussed effects due to the finite size of the DNA (line)
along which the diffusion takes place. This may be of importance for
certain in vitro experiments employing a rather short stretch of DNA.
The predicted behavior was corroborated (without adjustable parameter)
by simulations. Similar effects arise when the first passage of an
individual tagged particle is considered.

We note that our derivations were based on normal Markovian diffusion
dynamics. To generalize our results to situations footing on long-tailed
waiting time distributions, that cause a subdiffusive behavior, the
standard procedure can be used to map the Markovian to the subordinated
subdiffusive process \cite{report1}, and the associated dynamical equation
contains a fractional time derivative \cite{report,FraKi}. Intersegmental
jumps \cite{hippel} at places where by DNA-looping chemically distant
segments of the DNA get in close contact in physical three-dimensional
space \cite{looping}, can even give rise to L{\'e}vy flights
\cite{Paradox,JLum}.
The latter situation requires special care when comparing the efficiency
between sliding motion along the DNA and the L{\'e}vy flight mixing under
varied salt conditions, as explored in \cite{michael}. Another remark
concerning our modelling in terms of the diffusion-controlled Smoluchowski
picture is in order. Namely, in transport-controlled reactive systems, that
are not overdamped, at shorter times there is the need to include the
transient ballistic regime in the reaction scheme; as discussed in
Refs.~\cite{bere} starting from the Klein-Kramers picture. However, in
our problem, the diffusion process is highly overdamped \cite{hanggi} and the
Smoluchowski approach is therefore appropriate.

\begin{acknowledgments}
We would like to thank Ulrich Gerland and Oleg Krichevsky for helpful
discussions. This work was supported by NIH GM 072462, NSF MCB-0238190,
and the Research Corporation.
\end{acknowledgments}

\section{Appendix}

Here, we want to
elucidate the role of finite-size effects and to stress the difference
between the grand-canonical and the canonical situation (i.e., when
the number $N$ of particles in the interval of length $L$ is variable,
or fixed or prescribed by Poisson statistics, respectively). All our
considerations in the main text
were pertinent to the last situation corresponding to the grand-canonical
ensemble, which seems to be experimentally relevant.
Here, for completeness we discuss the other case.

We concentrate on the situation without adsorption-desorption processes
($k_0=k_1=0$). For noninteracting particles, the probability density $p(x,t)$
to find a particle at site $x$ is described by the same equation
(\ref{rde}), however
now with the initial condition $p(x,0)=1/L$ corresponding to the normalization
of the probability density. The overall survival probability of a given particle
in the interval is simply given by $\Psi(t)=\int_0^L p(x,t)dx$. Performing
integration over $x$ in equation (\ref{concentr}) giving now (for $n_0=1/L$,
$k_0=k_1=0$) the Laplace-transformed $\tilde{p}(x,u)$ 
\begin{equation}
\tilde{\Psi}(u)=\frac{1}{u}-\frac{\sqrt{D_{\mathrm{1d}}}}{L u^{3/2}} \tanh \left(\frac{L \sqrt{u}}{\sqrt{D_{\mathrm{1d}}}}\right).
\end{equation}
For exactly $N$ particles, the survival probability of the reaction center is 
$\mathscr{S}(t)=\Psi^N(t)$: it only survives if none of the particles arrived at it up
to the time $t$, and the mean survival time $T(N)=\int_0^\infty \mathscr{S}(t) dt =
\int_0^\infty \Psi^N(t)dt$. For $N=0$ one has $\mathscr{S}(t)=1$, so that the mean
waiting time diverges. For whatever finite $N$ the mean waiting time is finite.
It follows from the fact that the function $\Psi(t)$, which is non-negative and
monotonously non-growing, is integrable, and its time integral $T(1)=\lim_{u
\rightarrow 0} \tilde{\Psi}(u) = L^2/3D_{\mathrm{1d}}$. This means that for $t \rightarrow
\infty$ this function decays faster than as $t^{-1}$, and thus its powers decay
even faster, and are integrable. For $N$ small the value of $T(N)$ has to
be calculated explicitly. For large $N$ a simple asymptotic expression arises:
In this case the mean waiting time is much smaller than $L^2/D_{\mathrm{1d}}$, with small
times corresponding to large $u \gg D_{\mathrm{1d}}/L^2$. For such $u$ one has 
$\tanh \left[\sqrt{D_{\mathrm{1d}}} / \left( L \sqrt{u}\right) \right] \rightarrow 1$ so that
one can put down
\begin{equation}
\tilde{\Psi}(u) \simeq =\frac{1}{u}-\frac{\sqrt{D_{\mathrm{1d}}}}{L u^{3/2}}. 
\end{equation} 
The inverse Laplace transform of this function gives us the small-$t$ behavior
of $\mathscr{S}(T)$, namely 
\begin{equation}
\Psi(t) \simeq 1- \frac{1}{L} \sqrt{\frac{4D_{\mathrm{1d}}t}{\pi}}.
\label{shorttime}
\end{equation}
For $N$ large enough one then has
\begin{eqnarray}
\mathscr{S}(t)=\Psi^N(t) & \simeq & \left( 1- \frac{1}{L} \sqrt{\frac{4D_{\mathrm{1d}}t}{\pi}}
\right)^N \nonumber \\
& \simeq & \exp\left[N \ln \left( 1- \frac{1}{L} \sqrt{\frac{4D_{\mathrm{1d}}t}{\pi}}
\right)\right] \nonumber \\
& \approx & \exp \left(-\frac{N}{L} \sqrt{\frac{4D_{\mathrm{1d}}t}{\pi}}
\right),
\label{canonic}
\end{eqnarray}
which is exactly our equation (\ref{Pnoad}) with $n_0=N/L$. The approximation
is reasonably good starting from $N \sim 10$ particles. The canonical mean
waiting time (i.e., the mean waiting time with \textit{exactly} $N$ particles
in the interval) is then given by the same equation (\ref{tau}).

The grand-canonical result can be obtained from the canonical one by simply
noting that the grand-canonical expression for $\mathscr{S}(t)$ corresponds to a
weighted sum of the corresponding canonical waiting times:
\begin{equation}
\mathscr{S}(t)=\sum_N \Psi^N(t) p_N,
\end{equation} 
where $p_N$ is the probability to find exactly $N$ particles within the
interval. Taking this probabilities to follow a Poisson distribution, $p_N =
[(n_0L)^N/N!]\exp(-n_0L)$ we get:
\begin{eqnarray}
\mathscr{S}(t) & = & \sum_{N=0}^\infty \frac{\left[n_0 L \Psi(t)\right]^N}{N!}\exp(-n_0L)  \\
& = & \exp \left[ n_0L \left( \Psi(t)-1 \right) \right], \nonumber
\label{grandcan}
\end{eqnarray} 
which, for small $t$ again corresponds to our equation (\ref{Pnoad}). Using an
approximate short-time asymptotic expression  for $\Psi(t)$,
equation (\ref{shorttime}) we again arrive at equation (\ref{canonic}). 
However, some care is required when interpreting this result. 

Since  $\Psi(t)$ is integrable and thus $\Psi(t) \rightarrow 0$ for $t
\rightarrow \infty$, $\mathscr{S}(t)$ tends for $t \rightarrow \infty$ to a constant
value, $\mathscr{S}(t) \rightarrow \exp(-n_0 L)$, which is exactly the probability to
have no TFs in the interval. In this case, of course, no reaction takes place
at all. The mean waiting time, being the time-integral of $\mathscr{S}(t)$, clearly,
diverges. If we separate the first, constant, term in equation (\ref{grandcan}),
we get an expression for $\mathscr{S}(t)$ which (for $n_0L \gg 1$) is asymptotically the
same as equation (\ref{canonic}), and calculate the mean waiting time, we
again arrive at equation (\ref{tau}). This mean waiting time has however to
be interpreted as the mean waiting time for the reaction, provided that it
happens at all. The probability that it never happens is equal to $\exp(-n_0
L)$, and is small but finite.

\end{document}